\documentclass[aps,prl,twocolumn, notitlepage, superscriptaddress]{revtex4-1}

\usepackage[english]{babel}

\makeatletter
\adddialect\l@ENGLISH\l@english
\makeatother

\usepackage[utf8]{inputenc}
\usepackage{float}
\usepackage{graphicx}
\usepackage{amsmath}
\usepackage{amssymb}
\usepackage{xcolor}
\usepackage{physics}
\usepackage{upgreek}
\usepackage{xparse}
\usepackage{xr}

\newcommand{\NIST}{National Institute of Standards and Technology, Boulder, CO}
\newcommand{\CU}{Department of Physics, University of Colorado, Boulder, CO}

\newcommand{\ANL}{HEP Division, Argonne National Laboratory, Lemont, IL}

\newcommand{\ketda}{\ket{\downarrow}}
\newcommand{\ketua}{\ket{\uparrow}}
\newcommand{\brada}{\bra{\downarrow}}
\newcommand{\braua}{\bra{\uparrow}}

\newcommand{\Al}{{}^{27}\textrm{Al}^{+}}
\newcommand{\Ca}{{}^{40}\textrm{Ca}^{+}}
\newcommand{\Mg}{{}^{25}\textrm{Mg}^{+}}

\makeatletter
\newcommand*{\addFileDependency}[1]{
  \typeout{(#1)}
  \@addtofilelist{#1}
  \IfFileExists{#1}{}{\typeout{No file #1.}}
}
\makeatother

\newcommand*{\myexternaldocument}[1]{
    \externaldocument{#1}
    \addFileDependency{#1.tex}
    \addFileDependency{#1.aux}
}

\begin{document}
\raggedbottom
% not the best version of the title but want to get across that the clocks are separated, they are correlated, and the  we are probing at the lifetime limit or beyond the laser coherence time.  I like it (DBH)

\title{Lifetime-Limited Interrogation of Two Independent $\Al$ Clocks Using Correlation Spectroscopy}

%\title{Correlated Distant Atomic Clocks With Coherence at the Lifetime Limit}

% OTHER OPTIONS
% \title{Comparing distant atomic clocks beyond the local oscillator coherence time}
% \title{Using Correlations Between Separate Atomic Clocks to Surpass Laser Coherence Time}

%To Do: \phi_A to \phi_1?
% discuss timing errors again

\author{Ethan R. Clements}
\email{ethan.clements@nist.gov}
\affiliation{\NIST}
\affiliation{\CU}
\author{May E. Kim}
\affiliation{\NIST}
\author{Kaifeng Cui}
\affiliation{\NIST}
\affiliation{\ANL}
\author{Aaron M. Hankin}
\altaffiliation[Present address: ]{Honeywell Quantum Solutions, Broomfield, CO 80021}
\affiliation{\NIST}
\affiliation{\CU}
\author{Samuel M. Brewer}
\altaffiliation[Present address: ]{Colorado State University, Fort Collins, CO 80523}
\affiliation{\NIST}
\author{Jose Valencia}
\affiliation{\NIST}
\affiliation{\CU}
\author{Jwo-Sy A. Chen}
\altaffiliation[Present address: ]{IonQ Inc., College Park, MD 20740}
\affiliation{\NIST}
\affiliation{\CU}
\author{Chin-Wen Chou}
\affiliation{\NIST}
\author{David R. Leibrandt}
\affiliation{\NIST}
\affiliation{\CU}
\author{David B. Hume}
\email{david.hume@nist.gov}
\affiliation{\NIST}

\date{\today}

\begin{abstract}
Laser decoherence limits the stability of optical clocks by broadening the observable resonance linewidths and adding noise during the dead time between clock probes. Correlation spectroscopy avoids these limitations by measuring correlated atomic transitions between two ensembles, which provides a frequency difference measurement independent of laser noise. Here, we apply this technique to perform stability measurements between two independent clocks based on the $^1S_0\leftrightarrow{}^3P_0$ transition in $^{27}$Al$^+$. By stabilizing the dominant sources of differential phase noise between the two clocks, we observe coherence between them during synchronous Ramsey interrogations as long as 8 s at a frequency of $1.12\times10^{15}$~Hz. The observed contrast in the correlation spectroscopy signal is consistent with the 20.6~s $^3P_0$ state lifetime and supports a measurement instability of $(1.8\pm0.5)\times10^{-16}/\sqrt{\tau/\textrm{s}}$ for averaging periods longer than the probe duration when deadtime is negligible.
\end{abstract}

\maketitle

%best possible stability from lifetime limit = 1.4587240505718463e-16 at time = [10.3003003]

%\section{Introduction}
%Frequency comparisons
High-stability frequency comparisons are the basis of many applications of optical atomic clocks including time and frequency metrology~\cite{riehle_cipm_2018}, relativistic geodesy~\cite{mehlstaubler_atomic_2018}, and tests of fundamental physics~\cite{safronova_search_2018}. Measurements with optical clocks are typically performed by interrogating an atomic resonance using an ultrastable laser, and stabilizing the laser frequency based on the measured atomic transition probabilities~\cite{ludlow_optical_2015}. Here, laser frequency noise contributes intrinsically to measurement instability because it limits the probe duration~\cite{riis_optimum_2004,leroux_-line_2017}, effectively broadening the linewidth of the atomic resonance~\cite{itano_quantum_1993}. It also introduces noise during the dead time between clock interrogations~\cite{dick_local_1987}. Recent experiments have made improvements to the stability of laser systems but have yet to reach the stability required to probe many atomic clock transitions at the atomic species' natural linewidths~\cite{matei_1.5_2017}. Correlation spectroscopy is an alternative frequency comparison measurement technique that avoids these limitations by simultaneous interrogation of two atoms (or two atomic ensembles) with the same laser, which allows for common-mode cancellation of laser noise and probe times longer than the laser coherence time.

To illustrate the laser-noise limitation, consider frequency measurements on a two-level system with states $\ketda$ and $\ketua$~\cite{itano_quantum_1993}. A typical Ramsey sequence beginning from $\ketda$ involves two $\pi/2$-pulses with a controlled laser phase difference $\phi$ separated by the probe duration $T_\textrm{R}$~\cite{norman_f_ramsey_molecular_1956}. We assume each $\pi/2$ pulse has a duration negligible compared to $T_\textrm{R}$. A measurement of $\hat{\sigma}_\textrm{z} = \ketua\braua - \ketda\brada$ at the end of this sequence has expectation value $\langle\hat{\sigma}_\textrm{z}\rangle = \cos\left[\left(\omega_\textrm{L}-\omega_\textrm{0}\right)T_\textrm{R}+\phi\right]$, where $\omega_\textrm{L}$ is the laser frequency and $\omega_\textrm{0}$ is the atomic resonance frequency. Atom-laser decoherence (for example, due to laser frequency fluctuations or atomic spontaneous emission) alters this picture by reducing the contrast of the Ramsey fringe by a factor $C(T_\textrm{R})< 1$, which depends on the probe duration. In many optical clocks, including the $\Al$ clocks in this letter, decoherence over the relevant timescales is dominated by flicker-frequency noise of the laser~\cite{ludlow_optical_2015}. This limits the probe duration that minimizes measurement instability, which has been evaluated analytically and through numerical simulation~\cite{rosenband2009alpha,riis_optimum_2004,leroux_-line_2017,supplemental}. The reduced contrast $C(T_\textrm{R})$ due to flicker-frequency noise can be estimated based on the assumption of Gaussian-distributed phase fluctuations as $C(T_\textrm{R}) = e^{-(\sigma_0\omega_0 T_\textrm{R})^2/2}$, where $\sigma_0$ is the fractional flicker noise floor of the Allan deviation. The instability at long averaging times $\tau$ is then given by 
\begin{equation}
\label{eq:sigmaopt}
    \sigma({\tau}) = \frac{1}{\omega_\textrm{0} \sqrt{T_\textrm{R}\tau}}e^{(\sigma_\textrm{0}\omega_\textrm{0} T_\textrm{R})^2/2},
\end{equation}
which has a minimum at $T_\textrm{R} =1/\sqrt{2}\sigma_{\textrm{0}}\omega_\textrm{0}$. 
%As a simple model, if this decoherence is described by an exponential decay rate $\gamma$ (e.g. white frequency noise), then $C(T_R) = e^{-\gamma T_R}$ and the choice $T_R = 1/\gamma$ minimizes the variance in a frequency measurement.  In practice, the presence of flicker or random-walk frequency noise complicates this picture often resulting in $T_R$ significantly below the laser coherence time in order to optimize clock stability.

To avoid this limit, in correlation spectroscopy, two atoms or atomic ensembles are probed simultaneously with the same laser and their frequency difference is determined by measurements of the parity operator, $\hat{\Pi} = \hat{\sigma}_{\textrm{z},1}\otimes\hat{\sigma}_{\textrm{z},2}$. For unentangled atoms in a pure quantum state, ${\langle\hat{\Pi}\rangle = \langle\hat{\sigma}_{\textrm{z},1}\rangle\langle\hat{\sigma}_{\textrm{z},\textrm{2}}\rangle = \cos\left(\Delta_1 T_\textrm{R} + \phi_1\right)\cos\left(\Delta_2 T_\textrm{R} + \phi_2\right)}$, where we have defined $\Delta_i \equiv \omega_{\textrm{L}} - \omega_{0,i}$ and $i$ is an index that refers to each atom. We can separate $\langle\hat{\Pi}\rangle$ into terms that depend on the sum and difference frequency detunings $\Delta_{\pm} \equiv \Delta_1 \pm \Delta_2$ and phases $\phi_{\pm} \equiv \phi_1 \pm \phi_2$ such that
\begin{equation}
    \label{eq:CorrSpec1}
    \langle\hat{\Pi}\rangle = \frac12\left[\cos(\Delta_+ T_\textrm{R} + \phi_+) + \cos(\Delta_- T_\textrm{R} + \phi_-)\right].
\end{equation}
At probe durations long compared to the laser coherence time, the first term in Eq.~(\ref{eq:CorrSpec1}) averages to zero. The fundamental limit in coherence time for a particular clock transition is given by the spontaneous decay rate $\Gamma$ (typically the rate of decay from the excited state). If a spontaneous decay event occurs during the Ramsey probe duration the second Ramsey $\pi/2$-pulse places the atom in an equal superposition of up and down. Including spontaneous decay and assuming no laser coherence, Eq.~(\ref{eq:CorrSpec1}) becomes
\begin{equation}
    \label{eq:CorrSpecGamma}
    \langle\hat{\Pi}\rangle = \frac12e^{-\Gamma T_\textrm{R}}\cos\left(\Delta_- T_\textrm{R} + \phi_-\right).
\end{equation}
Since $\Delta_- = \omega_{0,1}-\omega_{0,2}$, Eq.~(\ref{eq:CorrSpecGamma}) represents a direct atom-atom frequency measurement that is independent of the laser noise. The fractional instability of a frequency ratio measurement at this lifetime limit is given by
\begin{equation}
\label{eq:sigma_corr}
    \sigma_{\rm D}(\tau) = \frac2{\omega_0\sqrt{T_{\rm R}\tau}}e^{\Gamma T_{\rm R}},
\end{equation}
where we have used $\omega_{0,i} \approx \omega_0$. The optimum probe duration for minimum instability of a correlation spectroscopy comparison is then $T_{\rm R,opt} = 1/(2\Gamma)$~\cite{supplemental}. 

Previous implementations of correlation spectroscopy for optical clocks used two or more ions~\cite{chwalla_precision_2007,olmschenk_manipulation_2007,chou_quantum_2011,tan_suppressing_2019,shaniv_quadrupole_2019,manovitz_precision_2019} or neutral atomic ensembles~\cite{2018_marti_imaging, young_tweezer_2020} confined in the same trap. In these experiments, the atoms were co-located to within a few microns such that differential effects including optical path length fluctuations and noise due to variations in the ambient electromagnetic field were naturally common-mode and thus suppressed. Using this technique for many clock applications requires implementation in spatially separated optical clocks where differential noise can limit their relative coherence. Here, by suppressing sources of differential noise, both in the probe laser beams and the atomic resonance frequencies, we demonstrate correlation spectroscopy between two independent clocks and observe linewidths approaching the ultimate limit of resolution from the $\Al$ ${^3\!P_0}$ excited-state lifetime of $20.6$~s~\cite{rosenband_observation_2007}.

%\section{Experiment} 
We implement correlation spectroscopy using two optical atomic clocks based on quantum logic spectroscopy of the $^1S_0 \leftrightarrow {}^3P_0$ transition in $\Al$. A key difference between the two optical clocks is the choice of qubit species, which is used for sympathetic cooling and state readout~\cite{schmidt_spectroscopy_2005}. One of these systems, using hyperfine levels in the ground state manifold of $\Mg$ as the qubit, has recently been evaluated to have a  systematic fractional frequency uncertainty of $\Delta f/f= 9.4\times10^{-19}$~\cite{brewer_+_2019}. The second, using the $S_{1/2}$ and $D_{5/2}$ levels of $\Ca$ as an optical qubit, is a newly-developed clock with improved control of some systematic uncertainties, but its error budget has not been fully evaluated. In what follows, we identify these two systems as $\Mg/\Al$ and $\Ca/\Al$, respectively.

\begin{figure}[t]
\includegraphics[scale = 0.65]{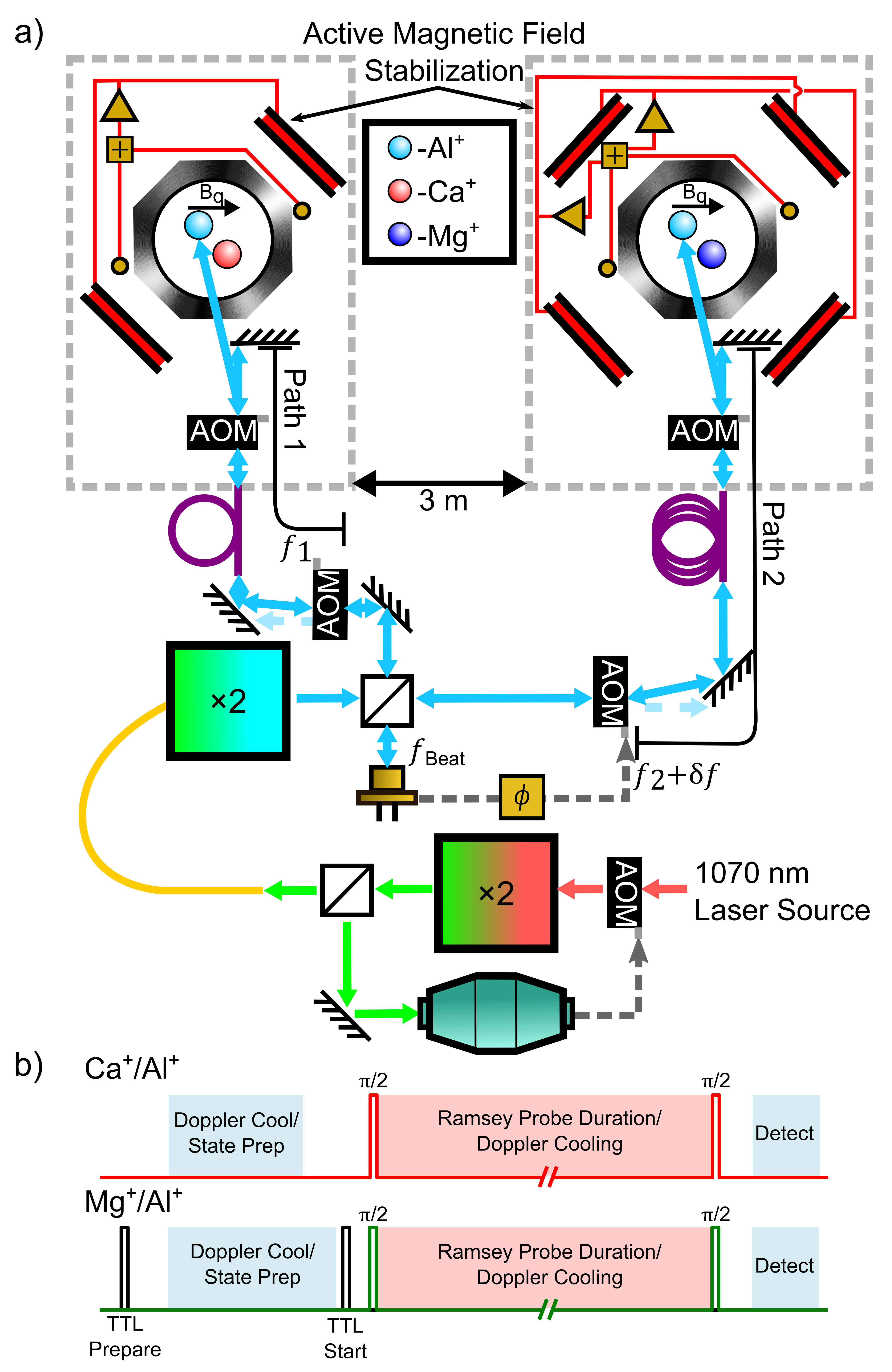}
\caption{a) Illustration of the correlation spectroscopy experiment, including simplified schematics of the laser path-length stabilization and active magnetic field stabilization setups. Here ${f_{\rm Beat} = 2(f_{1}-f_{2})}$ is phase locked to a maser-referenced 10 MHz signal, and the relative phase is corrected by modulating the Path 2 AOM, denoted by $f_2 + \delta f$. The magnetic field is stabilized using measurements from single-axis fluxgate sensors (shown as yellow circles) oriented along the quantization axis B$_{\rm{q}}$. In the $\Mg/\Al$ clock two pairs of coils are used, while in the $\Ca/\Al$ system there is only one. Boxes labeled $\times$2 denote frequency doubling of the input light where the final light sent to the atomic clocks is at 267.4~nm. b) Pulse sequence for the synchronized interrogation. Line breaks indicate that the clock interrogation is much longer than the detection, cooling, and state preparation required on each cycle of the sequence.}
\label{fig:experimental_setup}
\end{figure}

The two clocks are located on optical tables spaced roughly 3~m apart. A diagram of the experiment is given in Fig.~\ref{fig:experimental_setup}(a). All laser systems used for cooling and manipulation of the qubit ions are independent; however, the $\Al$ laser systems ($^3P_1$ and $^3P_0$) both share a common source for the two clocks. The 267~nm laser light used to drive the $^1S_0\leftrightarrow{^3P_0}$ clock transition is generated on the $\Ca/\Al$ optical table and sent to the $\Mg/\Al$ table via a 6-m-long UV-cured photonic crystal fiber~\cite{colombe_single-mode_2014}. 

% c) Partial energy level diagram of the $\ket{^1S_0}$ and $\ket{^3P_0}$ Zeeman manifolds. The arrows denote the two optical clock transitions used here.
%\aspectratio*{main_text_figures/timing_setup_combined.png}[\test]
%\texttt{\meaning\test}
%\aspectratio*{main_text_figures/combined_fringe_fig.png}[\testa]
%\texttt{\meaning\testa}
%\aspectratio*{main_text_figures/combined_stability_contrast_vs_time.png}[%\testb]
%\texttt{\meaning\testb}

Using the same laser source for the two clocks allows for precise control of the differential phase in the probe pulses by active suppression of Doppler-noise in the optical fibers and free-space optical paths~\cite{ma_delivering_1994,ye_delivery_2003}. A diagram of the path-length stabilization setup is given in Fig.~\ref{fig:experimental_setup}(a), where the total path length between the two ions is $\approx$10~m. Part of the laser beams are retro-reflected close to where they enter the two vacuum systems and form a beatnote at a beamsplitter close to the UV frequency doubler. The relative phase noise in this beatnote is measured using a 400~MHz bandwidth avalanche photodiode and is stabilized by controlling an acousto-optic modulator (AOM) frequency in the $\Mg/\Al$ path. In out-of-loop measurements using a test setup comparable to the setup in Fig.~\ref{fig:experimental_setup}, we observe differential phase fluctuations below $\pi/20$ at Ramsey probe durations as long as 12~s~\cite{supplemental}. This residual noise is likely limited by the short, out-of-loop, open-air paths such as those before the ion traps. When running the experiment, a frequency counter monitors the in-loop beat-note to check for cycle slips in the phase-locked loop.

\begin{figure*}[t]
\includegraphics[scale = 0.95]{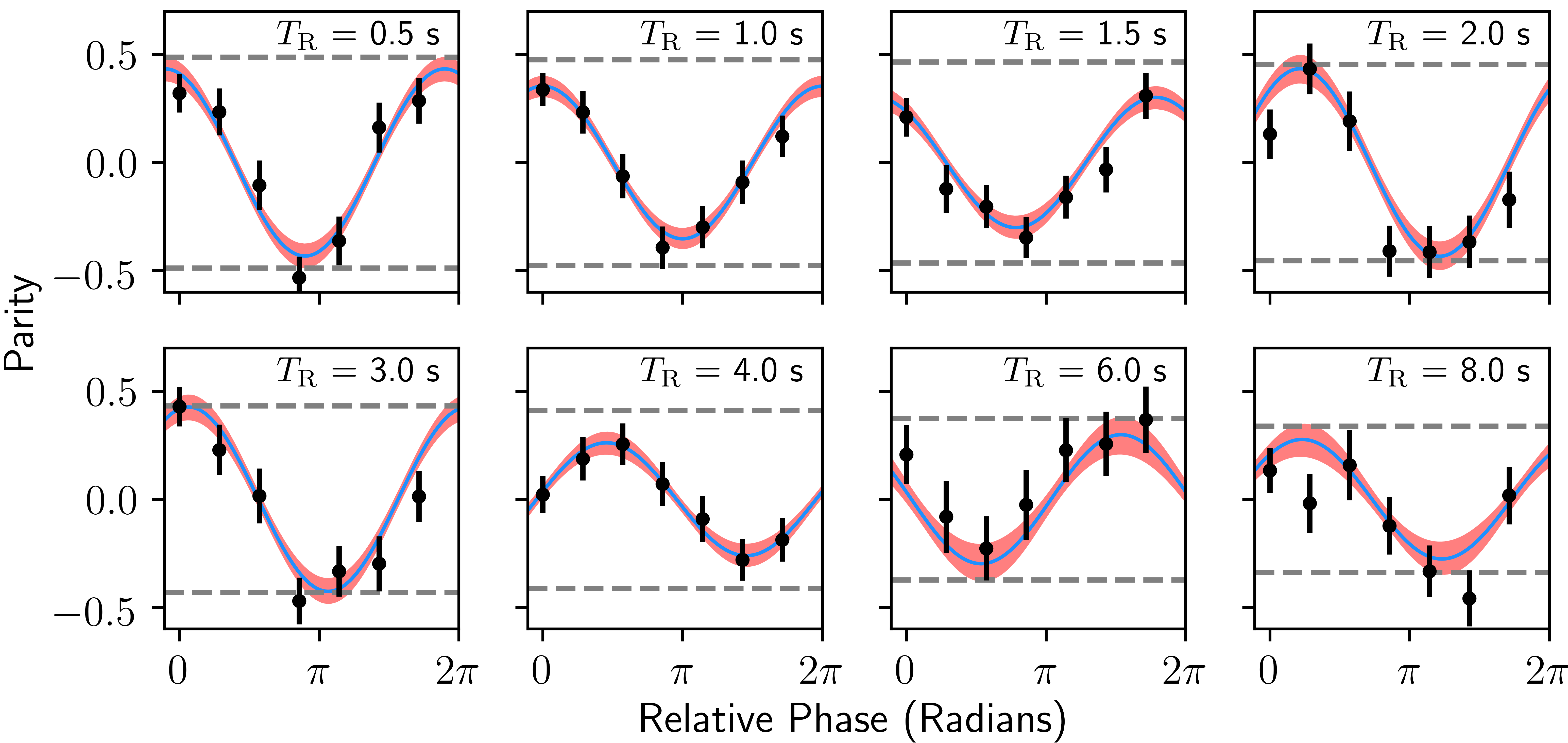}
\caption{\label{fig:combined_fringe} Parity fringes obtained for Ramsey probe durations between 0.5~s and 8~s (upper right labels). Here, the transition used for correlation spectroscopy is the $\ket{^1S_0,m_F=3/2}~\leftrightarrow~\ket{^3P_0,m_F=1/2}$ transition. Experimental data (black dots) are shown with error bars dominated by quantum projection noise. Fits to these parity fringes (blue lines) and their 1$\sigma$ confidence intervals (red shading) are determined by re-sampling the data using non-parametric bootstrapping methods. The maximum obtainable parity amplitude (gray dashed lines) due to the finite lifetime of the two $\Al$ ions is calculated using Eq.~(\ref{eq:CorrSpecGamma}).
}
\end{figure*}

Another effect that can limit the atom-atom coherence of the two systems is fluctuations of the local magnetic fields. To minimize the corresponding Zeeman shifts in each clock, we servo the magnetic field based on measurements with multiple fluxgate magnetometers placed close to the vacuum chamber and oriented along the clock quantization axis. A linear combination of these measurements is used to estimate the magnetic field at the ion and corrections are made using a set of Helmholtz coils mounted around each optical table. Using these active stabilization techniques, we reduce the magnetic field noise amplitude to below 20~${\upmu}$Gauss (${\upmu}$G) for averaging times as long as $10^3$~s~\cite{supplemental}. 

Both the $^1S_0$ and $^3P_0$ states in $\Al$ have a magnetic quantum number of $F = 5/2$. We performed initial correlation spectroscopy experiments on the $\ket{^1S_0,m_F=5/2}~\leftrightarrow~\ket{^3P_0,m_F=5/2}$ transition, which has a sensitivity to magnetic fields of $-4.2$~kHz/G (1~G = $10^{-4}$~T). Through numerical simulations using measured magnetic field noise, we found that this residual magnetic field noise was still a limitation~\cite{supplemental}. To further reduce the effect of magnetic field noise, we switched to probing the $\ket{^1S_0, m_F = 3/2}\leftrightarrow\ket{^3P_0,m_F = 1/2}$ transition. This transition has a sensitivity to magnetic fields of 0.28~kHz/G, a factor of $\approx$15 reduction in sensitivity compared to the typical clock transition. Preparation of the $\ket{^1S_0,m_F = 3/2}$ initial state is done by applying a series of $\pi$-polarized laser pulses on the $\ket{{}^1S_0,m_{F} = m}\rightarrow\ket{{}^3P_1,F = 7/2,m_{F} = m}$ transitions, for $m \in \{{-5/2,-3/2,-1/2,1/2,5/2}\}$. The frequencies of the sequentially applied laser pulses are tuned to be on resonance for each $m_{F}$ transition. Because the ${}^3P_1,F=7/2$ manifold has $g_F\approx 3/7$~\cite{schmidt_spectroscopy_2005,guggemos_frequency_2019}, the splitting of the transition frequencies for adjacent Zeeman levels is near 1~MHz at typical operating magnetic fields of 1.5 to 1.7~G, and these optical pumping transitions (pulse durations $t_{\pi} > 50$~$\rm{\mu s}$) are frequency-resolved. The target state is thus a dark state of the optical pumping process. The series of $\pi$-polarized laser pulses is repeated twelve times to ensure a high fidelity of being in the target state~\cite{supplemental}. Once $\ket{^1S_0, m_F = 3/2}$ is prepared, we drive the $\ket{^1S_0, m_F = 3/2}\leftrightarrow\ket{^3P_0,m_F = 1/2}$ transition with a $\sigma^+/\sigma^-$-polarized laser beam.

Synchronization between the two experimental control systems is achieved in a transmit/receive configuration. The $\Mg/\Al$ system takes the role of the transmitter and supplies the $\Ca/\Al$ system with triggering pulses; the experimental sequence can be seen in Fig.~\ref{fig:experimental_setup}(b). To begin the experiment, the $\Mg/\Al$ system sends a ``prepare'' TTL pulse to the $\Ca/\Al$ system, which initiates the laser cooling and state preparation sequences required before interrogating the clock transition. The $\Ca/\Al$ system (which requires less time for preparation) then waits for a ``start" TTL indicating that the $\Mg/\Al$ system is finished with its cooling and state preparation. After the ``start" TTL each clock waits for a (different) predefined time, which is used to manually account for a constant communication lag between the two systems. Subsequently, the two systems drive the first of the two $\pi/2$ Ramsey pulses on the corresponding atomic clocks. The clocks' states evolve for the Ramsey period $T_\textrm{R}$, with continuous sympathetic Doppler cooling applied to the qubit ion~\cite{rosenband_observation_2007}. Following the Ramsey probe duration the second $\pi/2$ pulse is applied to each clock. The relative phase of the second $\pi/2$ pulse between the two systems is scanned. Finally, the state of the atom is measured using quantum-logic-based readout and recorded for post-processing calculations of the parity. 

During a measurement run, we use the measurement outcome of the previous experimental cycle as projective state preparation for the next such that $\ketda$ can be either the $^1\!S_0$ or $^3\!P_0$ state. Parity measurements are made by observing if a transition in each ion state has occurred since the previous interrogation. A parity of +1 corresponds to both atoms making a transition or both not making a transition, whereas a parity of $-1$ corresponds to only one of the two ions making a transition. To generate the parity fringes seen in Fig.~\ref{fig:combined_fringe}, the $\Ca/\Al$ clock is interrogated with a constant Ramsey phase $\phi_1$, while the $\Mg/\Al$ clock scans its phase $\phi_2$, relative to the $\Ca/\Al$ clock. By scanning the relative phase between the two systems' second $\pi/2$-pulses, $\phi_-$ can be scanned allowing the coherence between the two systems to be observed. Each point on the correlation spectroscopy fringe is probed $\gtrsim$50 times to average down the quantum projection noise.

In these parity phase scans, we observe atom-atom coherence well beyond the coherence time of the laser ($460\pm30$~ms), which has been measured using a single ion~\cite{supplemental}. Due to periodic interruptions from ion loss and other effects, which are filtered from the data as described in the supplement~\cite{supplemental}, the fringes in Fig.~\ref{fig:combined_fringe} accumulate data from multiple runs of the experiment and span total measurement durations as long as 4 hours. The fringe contrast thus represents all atom-atom decoherence mechanisms that act on timescales of seconds as well as long-term frequency drifts that act on timescales of hours. To maintain the laser frequency near resonance for the Ramsey $\pi/2$ pulses between these runs, common-mode adjustments to the laser frequency were made. 

\begin{figure*}
    \centering
    \includegraphics[scale = 0.95]{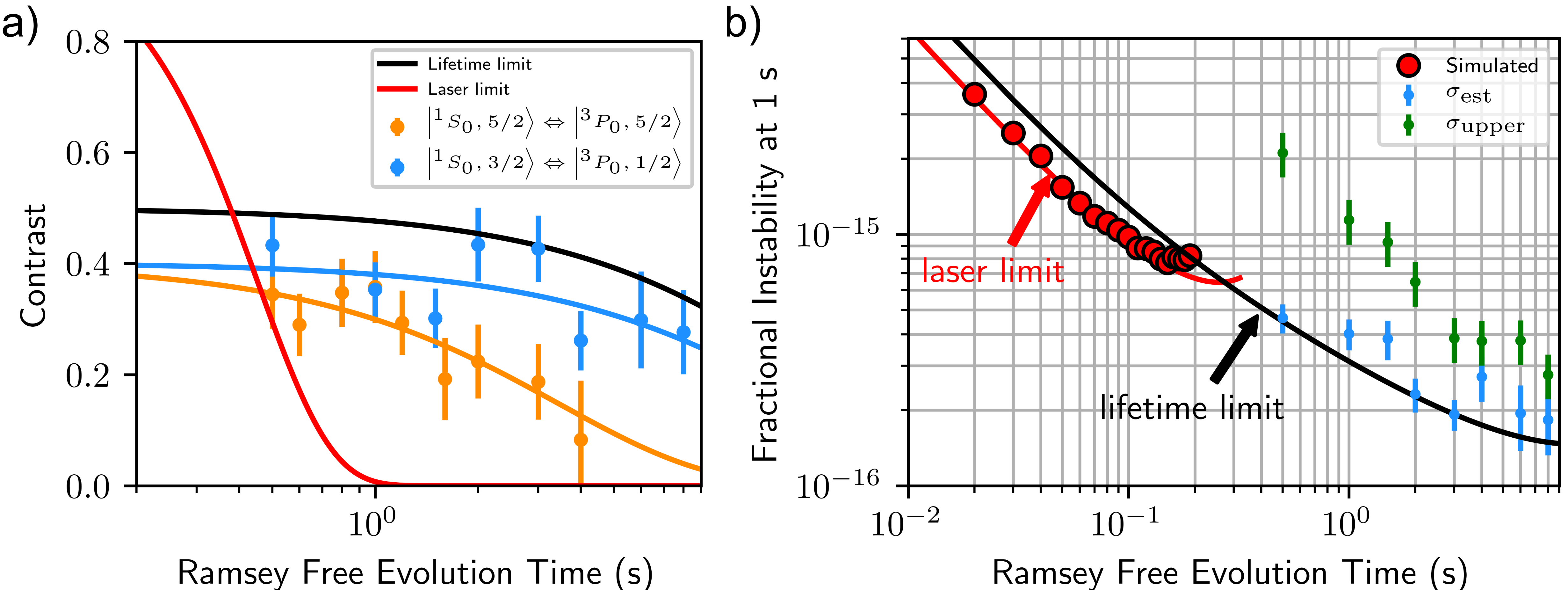}
    \caption{\label{fig:con_stab_v_time} a) Contrast as a function of the probe duration. The measured contrast (solid points) and associated uncertainty come from fits to the parity fringes. For comparison, a fit to the laser-coherence-limited Ramsey spectroscopy contrast~\cite{supplemental} (red line) and the calculated upper bound on the correlation spectroscopy contrast set by the lifetime limit (black line) are plotted. A fit to the experimental points using the model function $A\exp^{-T_\textrm{R}/t_\textrm{d}}$ is determined, where $A$ is the contrast and $t_\textrm{d}$ is the decay time. Fitting with this function gives A = $0.4 \pm 0.04$ and $t_\textrm{d} = 19 \pm 11$~s for $\ket{^1S_0,m_F=3/2}\leftrightarrow\ket{^3P_0,m_F=1/2}$ and A = $0.4 \pm 0.06$ and $t_\textrm{d} = 4 \pm 2$~s for $\ket{^1S_0,m_F=5/2}\leftrightarrow\ket{^3P_0,m_F=5/2}$. b) Comparison of the instability calculations and measurements as a function of probe duration. The instability $\sigma_{\rm upper}$, calculated using Eq.~(\ref{eq:sigmaupper}), is shown with green dots. This can be compared against the instability $\sigma_{\rm est}$ determined with Eq.~(\ref{eq:sigmaest}) shown with blue dots. A lower bound on the instability is given by the lifetime limit (black line, Eq.~(\ref{eq:sigma_corr})), which assumes a randomized laser phase at all probe durations. Also included is an estimate of the instability at the laser-noise limit both from the analytical estimate (red line, Eq.~(\ref{eq:sigmaopt})) and a numerical simulation (red points) assuming a flicker frequency noise floor at $4.4\times10^{-16}$. Numerical simulations stop at a probe duration of $\approx~200$~ms due to fringe hops occurring in our numerical simulation. For all theoretical estimates we assume a dead time of only 0.1~s (the average single-cycle dead time of our clocks), which has negligible impact at longer probe durations.}
\end{figure*}

%When using correlation spectroscopy to compare the two atomic clocks, the correlation signal can be converted into an error signal to lock the relative frequency of the two clocks. 
%This measure of the relative frequency would then be used to evaluate the accuracy of two systems.
%The data contributing to Fig.~\ref{fig:combined_fringe} have been filtered for several effects, all of which are monitored in real time. These include synchronization errors, collisions with background gas molecules~\cite{hankin_systematic_2019}, and events in which at least one of the $\Al$ ions is excited, via collisions, to a metastable internal state that is not addressable by laser pulses.  The likely metastable states for $\Al$ are $\ket{^3P_2}$ ($\tau=298.5$~s) and any Zeeman sublevel of $\ket{^3P_0}$ that is not addressed by the clock probe. The total percentage of useful data was as small as 24$\%$ for short $T_\textrm{R}$ and as large as 71$\%$ for longer $T_\textrm{R}$. Filtering of the data significantly reduces the mean duty-cycle for probing the clock transitions, but improves the contrast of the parity signal. 
%Discussion of Results
Fits of the function $\langle\hat{\Pi}(\phi_-)\rangle = C\cos(\phi_--\phi_0)$ to the parity data in Fig.~\ref{fig:combined_fringe} are used to extract the contrast $C$, phase $\phi_0$, and their associated uncertainties. The uncertainties are obtained by a bootstrapping method which resamples the experimentally determined binomial distributions~\cite{efron_bootstrap_1992, supplemental}. A plot of the measured contrast as a function of the Ramsey probe duration can be seen in Fig.~\ref{fig:con_stab_v_time}, showing data taken on the less magnetically sensitive $\ket{^1S_0,m_F = 3/2}\leftrightarrow\ket{^3P_0,m_F = 1/2}$ transition as well as initial data taken on the $\ket{^1S_0, m_F = 5/2}\leftrightarrow\ket{^3P_0, m_F = 5/2}$ transition. The noise suppression due to the magnetic field servo is comparable in both of these data sets and the improvement in the contrast is due to the reduced magnetic sensitivity of the $\ket{^1S_0,m_F=3/2}\leftrightarrow\ket{^3P_0,m_F=1/2}$ transition. 

The decay time of the contrast for experiments on the $\ket{^1S_0,m_F=3/2}\leftrightarrow\ket{^3P_0,m_F=1/2}$ transition is measured to be $t_\textrm{d} = 19 \pm 11$~s. This value is much longer than the measured laser coherence time of 460 $\pm$ 30~ms and is consistent with the decay time of $20.6$~s expected due to the finite excited-state lifetime. However, we observe a $20(8)\%$ reduction in the contrast from the ideal value of 0.5 set by Eq.~(\ref{eq:CorrSpecGamma}). We attribute this primarily to errors in the $\Al$ state preparation and $\pi$-pulse infidelity when driving the clock transition.

% To supplement
%  Each point along the fringe is re-sampled 10000 times to determine a distribution of possible results attainable given the experimentally measured binomial distribution. By fitting to each of these 10000 iterations of the re-sampled fringe, a distribution of possible phase offsets and contrast values is obtained. The distribution of the acquired fit parameters is also used to determine the uncertainty in the fit contrast and the phase offset.
The contrast of the fringes can be used to estimate the measurement instability for correlation spectroscopy comparisons between the two clocks~\cite{chou_quantum_2011}, using
\begin{equation}
\label{eq:sigmaest}
    \sigma_{est} = \frac{1}{\omega_0 C(T_\textrm{R})\sqrt{T_\textrm{R}}}.
\end{equation}
We find instability as low as $\sigma_{\textrm{est}} = (1.8\pm0.5)\times10^{-16}/\sqrt{\tau/\textrm{s}}$ at $T_{\rm R} = 8$~s, which corresponds to the achievable instability given the observed contrast if there is no dead time in the measurement and all probes were made at the relative phases where the parity slope is the highest. In our experiment, for the longer probe durations, we have negligible overhead due to state preparation and measurement, but suffer from frequent interruptions due to collisions with background gas. An upper estimate of the achievable measurement instability assumes a total averaging time $\tau_{\rm tot}$ including all dead time during the measurement runs, and the phase uncertainty $\sigma_{\phi}$ determined from the fit of the parity fringe,
\begin{equation}
\label{eq:sigmaupper}
    \sigma_{\rm upper} = \frac{\sigma_{\phi}\sqrt{\tau_{\rm tot}}}{T_\textrm{R} \omega_0}.
\end{equation}
This gives the measurement instability achieved in the phase scans presented in this letter, which is as low as $(2.8\pm0.6)\times10^{-16}/\sqrt{\tau/\textrm{s}}$ for $T_{\rm R} = 8$~s, as shown in Fig~\ref{fig:con_stab_v_time}. 

%If the hydrogen partial pressure in the $\Mg/\Al$ system were reduced to the level of the $\Ca/\Al$ system the achievable stability would be closer to value attributed to a measurement with no dead time.
%\section{Conclusion}
%Conclusion
In summary, we have demonstrated atomic coherence at probe durations as long as 8~s between optical resonances of two $\Al$ ions held in separate traps. The contrast $1/e$ decay time of $t_\textrm{d} = 19 \pm 11$~s is consistent with the 20.6 s excited state lifetime (corresponding to $2.3\times10^{16}$ optical cycles). Coherence at this level is sufficient to reach a ratio measurement instability below $3\times10^{-16}/\sqrt{\tau/\textrm{s}}$ for averaging times $\tau \gg T_\textrm{R}$. This stability supports a relative frequency measurement with statistical uncertainty $1\times10^{-18}$ in a single day of averaging.

Correlation spectroscopy between spatially separate atomic clocks could improve measurement precision for many applications of optical clocks in which a direct atom-atom comparison is needed. For example, relativistic geodesy measures the gravitational potential difference between two geographical locations by observing a relative frequency shift between atoms located at those points~\cite{mehlstaubler_atomic_2018}. This has been proposed as an alternative to existing geodetic survey techniques with potential advantages in terms of spatial and temporal resolution. By extending the probe duration beyond the laser coherence limit, future geodetic surveys could use portable laser systems with relatively poor stability compared to the best laboratory systems, but still average quickly to the limits imposed by clock accuracy. Similarly, extensions of this technique~\cite{hume_probing_2016,dorscher_dynamic_2019} to optical clocks based on different atomic species could be used to measure or constrain the time-variation of fundamental constants and to search for ultralight dark matter~\cite{safronova_search_2018}. These searches could achieve greater resolution by avoiding laser noise limits. Correlation spectroscopy takes advantage of the fact that atomic resonances can have a longer coherence time than that of the most stable laser demonstrated to date. It allows for the realization of many promising applications of optical clocks independent of further development of ultrastable laser technologies. 
\begin{acknowledgements}
We thank Daniel Cole and Raghavendra Srinivas for their careful reading and feedback on this manuscript. This work was supported by the NIST, DARPA, and ONR (Grant No. N00014-18-1-2634). M.E.K. was supported by by an appointment to the Intelligence Community Postdoctoral Research Fellowship Program at NIST administered by ORISE through an interagency agreement between the DOE and ODNI. S. M. B. was supported by ARO through MURI Grant No. W911NF-11- 1-0400. The views, opinions and/or findings expressed are those of the authors and should not be interpreted as representing the official views or policies of the Department of Defense or the U.S. Government.
\end{acknowledgements}

%\bibliography{CorrelationSpectroscopy40820}
\bibliography{CorrelationSpectroscopy052020}
\end{document}

\raggedbottom
\title{\textit{Supplemental Material for}\\
Lifetime-Limited Interrogation of Two Independent $\Al$ Clocks Using Correlation Spectroscopy}

\author{Ethan R. Clements}
\email{ethan.clements@nist.gov}
\affiliation{\NIST}
\affiliation{\CU}
\author{May E. Kim}
\affiliation{\NIST}
\author{Kaifeng Cui}
\affiliation{\NIST}
\affiliation{\ANL}
\author{Aaron M. Hankin}
\altaffiliation[Present address: ]{Honeywell Quantum Solutions, Broomfield, CO 80021}
\affiliation{\NIST}
\affiliation{\CU}
\author{Samuel M. Brewer}
\altaffiliation[Present address: ]{Colorado State University, Fort Collins, CO 80523}
\affiliation{\NIST}
\author{Jose Valencia}
\affiliation{\NIST}
\affiliation{\CU}
\author{Jwo-Sy A. Chen}
\altaffiliation[Present address: ]{IonQ Inc., College Park, MD 20740}
\affiliation{\NIST}
\affiliation{\CU}
\author{Chin-Wen Chou}
\affiliation{\NIST}
\author{David R. Leibrandt}
\affiliation{\NIST}
\affiliation{\CU}
\author{David B. Hume}
\email{david.hume@nist.gov}
\affiliation{\NIST}
\maketitle

%The supplementary information is organized as follows: In Sec.~\ref{sec:laser_noise} modeling and measurement of the laser noise and the effect it has on the measurement stability is presented. In Sec.~\ref{sec:correlation_stability} details are provided on the limits to measurement stability in correlation spectroscopy. In Sec.~\ref{sec:ca_al_clock} information on efforts to reduce background gas collisions in the new $\Ca/\Al$ clock other differences with the older $\Mg/\Al$ system are given. In Sec.~\ref{sec:optical_pumping} more detail is presented on the pumping scheme to prepare the $\Al$ ion in an arbitrary Zeeman level in the $^1S_0$ manifold. In Sec.~\ref{sec:differential} characterization of differential noise sources present in the experiment are detailed. In Sec.~\ref{sec:data_analysis} a brief discussion of bootstrapping techniques used in error analysis is made and suspected limitations to the parity amplitude are specified.

\section{Laser noise in Ramsey spectroscopy}
\label{sec:laser_noise}

In a typical clock comparison, laser coherence limits the resolution of frequency measurements on each clock individually. The dominant source of laser noise is often flicker frequency noise. Here, we model the effect of flicker frequency noise and compare this model with data resulting from Ramsey interrogation where the free evolution time is scanned.

\subsection{Analytical model}

The expectation value for $\hat{\sigma}_z$ at the end of a Ramsey experiment can be written as,
\begin{equation}
\label{eq:ramsey}
\tag{S1}
    \langle\hat{\sigma}_z\rangle = \cos\left[\left(\omega_{\rm L}-\omega_0\right)T_{\rm R}+\phi_{\rm N}+\phi\right],
\end{equation}
where $\omega_{\rm L}/2\pi$ is the laser frequency, $\omega_0/2\pi$ is the atom frequency, $T_{\rm R}$ is the Ramsey probe time, $\phi$ is the controlled laser phase difference between the first and second $\pi/2$ pulses and $\phi_N$ accounts for noise in the laser at the timescale $T_{\rm R}$.  A simple lower bound on the clock instability can be obtained by assuming negligible fluctuations in the atomic frequency $\omega_0$ and laser phase noise $\phi_{\rm  N}$ described by a Gaussian distribution (see Ref.~\cite{leroux_-line_2017}),  
\begin{equation}\label{eq:pgauss}\tag{S2}
P(\phi_{\rm N}) = \frac1{\sigma_{\rm N}\sqrt{2\pi}}e^{-\phi_{\rm N}^2/2\sigma_{\rm N}^2}.
\end{equation}
We assume slow feedback is used to correct for drifts in $\omega_L$ such that the flicker-noise limited $\phi_{\rm N}$ has a standard deviation $\sigma_{\rm N} = \sigma_0\omega_{\rm L} T_{\rm R}$, where $\sigma_0$ is the fractional flicker noise floor of the Allan deviation. Averaging Eq.~(\ref{eq:ramsey}) over this classical noise, we get
\begin{align}\label{eq:ramsey_decoherence}\tag{S3}
\displaystyle
\langle\hat{\sigma}_z\rangle =& \int_{-\infty}^{\infty}P({\phi_{\rm N}})\cos\left[\left(\omega_{\rm L}-\omega_0\right)T_{\rm R}+\phi_N+\phi\right]d\phi_{\rm N}\nonumber\\
\tag{S4}=& e^{-\sigma_{\rm N}^2/2}\cos\left[\left(\omega_{\rm L}-\omega_0\right)T_{\rm R}+\phi\right].
\end{align}
Therefore, laser noise reduces the contrast of the Ramsey fringe by the factor $C(T_{\rm R}) = e^{-(\sigma_0\omega_{\rm L} T_{\rm R})^2/2}$ and increases the single-shot measurement uncertainty ${\delta\omega_{\rm L} = \delta\hat{\sigma}_z/|d\langle\hat{\sigma}_z\rangle/d\omega_{\rm L}|}$~\cite{itano_quantum_1993}. For $\omega_L = \omega_0$, the choice $\phi = \pi/2$ maximises the error-signal slope \begin{equation}\label{eq:dsigmadz}\tag{S5}
\left|\frac{d\langle\hat{\sigma}_z\rangle}{d\omega_{\rm L}}\right| = T_{\rm R} e^{-(\sigma_0\omega_{\rm L} T_{\rm R})^2/2},
\end{equation}
and the projection-noise-limited frequency instability is given by
\begin{align}\label{eq:deltaomega_TR}\tag{S6}
\sigma_{y}(\tau) &= \frac{\delta\omega_{\rm L}}{\omega_0}\sqrt{\frac{T_{\rm R}}{\tau}},\\
\tag{S7}&= \frac{1}{\omega_0\sqrt{T_{\rm R} \tau}} e^{(\sigma_0\omega_0 T_{\rm R})^2/2}.
\end{align}
The optimum probe time in this model can be found by minimizing with respect to $T_{\rm R}$, giving
\begin{equation}\label{eq:TRopt}\tag{S8}
T_{\rm R,opt} = \frac{1}{\sqrt2 \sigma_0\omega_0}.
\end{equation}

%The initial state is $\ketda$ and can be either the ground or excited atomic state since the change is the atomic state is the measurement of interest. Laser noise limits the maximum probe time (inversely, the frequency resolution of the Ramsey experiment) and hence the clock instability.

 %On timescales from 0.1 s to 10 s, the noise in cavity stabilized lasers is often dominated by thermomechanical noise with a characteristic $1/f$ frequency noise spectrum.  Under these conditions, the fractional frequency noise is independent of averaging time and can be described by a single stability parameter $\sigma_0 = \delta\omega_L/\omega_L$.  The optimum Ramsey probe duration in the presence of this noise has been studied numerically and analytically~\cite{riis_optimum_2004,rosenband_alpha-dot_2009,leroux_-line_2017}, and it depends on the details of the laser noise spectrum, including temporal correlations.

This simple treatment gives a value close to the asymptotic optimum probe time reported in Ref.~\cite{leroux_-line_2017}. As described there, a more realistic treatment of the laser noise forces the maximum probe time to be shorter than $T_{\rm R,opt}$ to avoid Ramsey fringe hops. Below, we describe a measurement of the laser coherence time then use that measurement in a numerical model to take into account this more stringent limitation.

\subsection{Laser coherence measurement}

%To measure the coherence time of the laser (local oscillator), the frequency of the local oscillator is first de-tuned from resonance by 13 Hz and the ramsey free evolution time is scanned. From scanning the Ramsey free evolution time we expect to see an oscillation from the detuning with an envelope that is decaying due to the flicker frequency noise of the the laser. On timescales limited by the laser coherence, the variance in the laser frequency is independent of the Ramsey free evolution time and is given by $\sigma_0 = \delta\omega_L/\omega_L$. The distribution of the relative phase noise between the two $\pi/2$ pulses of the Ramsey sequence can be approximated as a gaussian distribution given by 
%\begin{equation}
%   P(\phi_L) = \frac{1}{\sigma_{\phi_L}\sqrt{2\pi}}e^{-\phi_L^2/2\sigma_{\phi_L}^2}
%\end{equation}
%with $\phi_L(T_R) = (\omega_L-\omega_0)T_R$ that has a standard deviation of $\sigma_{\phi_L}=\sigma_0\omega_LT_R$. Averaging over many scans of $T_R$ where this laser noise is present, we obtain an expression that is dependent on the laser phase noise and is given by
%\begin{equation}
%    \langle\hat{\sigma}_Z\rangle =e^{-(\sigma_0\omega_LT_R)^2/2}\cos({(\omega_L-\omega_0)T_R+\phi}).
%\end{equation}
\begin{figure}[h!]
\centering
\includegraphics[scale = 0.9]{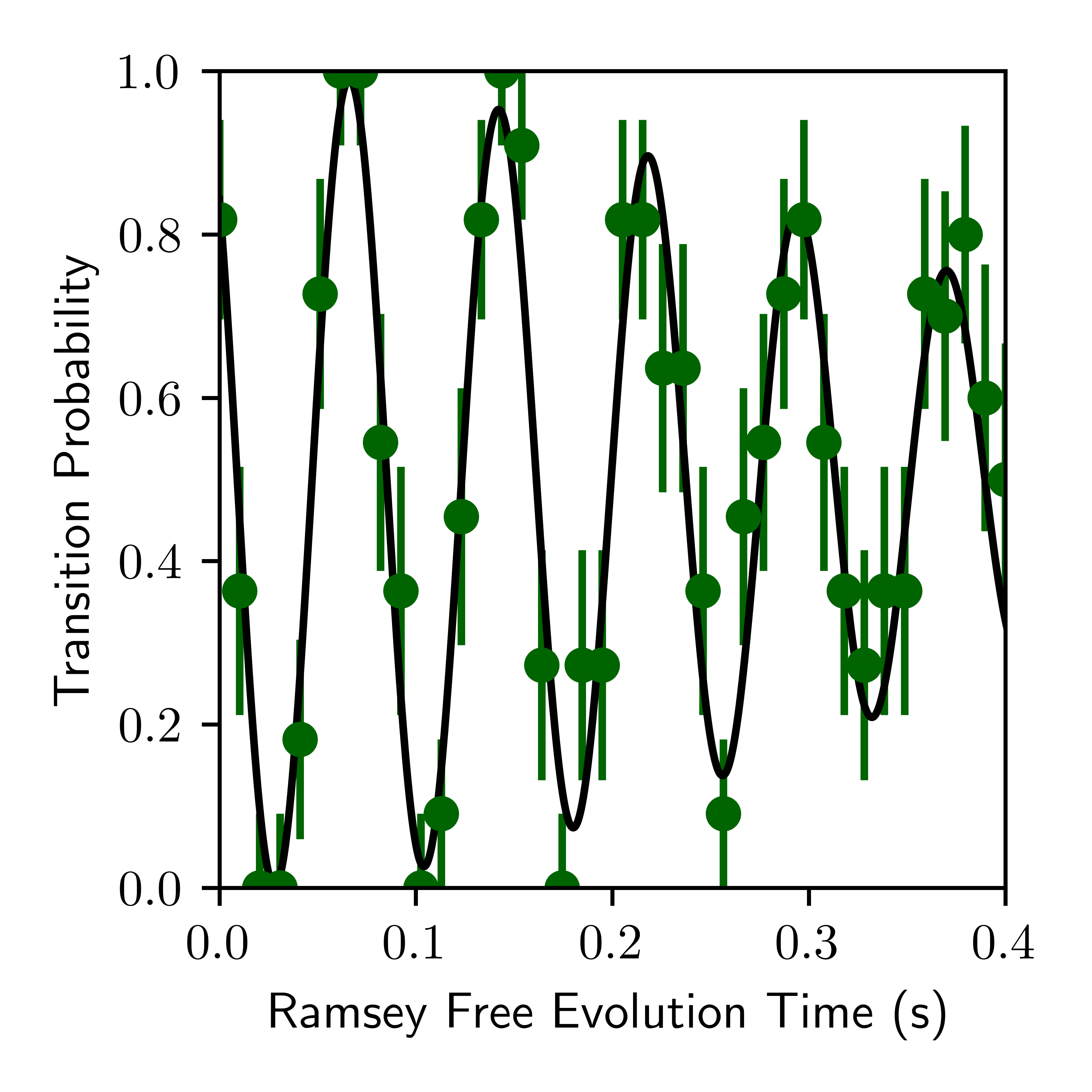}
\caption{\label{fig:laser_cohere} Plot of the transition probability as a function of the Ramsey free evolution time. The decay time of the envelope is set by the local oscillator noise. From fitting to the data we obtain a fractional laser instability $\sigma_0 = 4.4(2)\times10^{-16}/\sqrt{\tau/s}$.}
\end{figure}

To estimate our laser noise floor in the framework of this model, we performed a Ramsey experiment detuned by a known frequency $(\omega_{\rm L} - \omega_0)/2\pi = 13.14$~Hz. and fit to the resulting oscillations. Fitting Eq.~(\ref{eq:ramsey_decoherence}) to the resulting oscillations as seen in Fig.~S\ref{fig:laser_cohere} we obtain a fractional noise floor of $\sigma_0 = 4.4(2)\times10^{-16}/\sqrt{\tau/s}$. This corresponds to $T_{\rm R,opt} = 0.22$~s, roughly consistent with our typical (Rabi) probe time of 0.15 s.

\subsection{Simulated stability with flicker frequency noise}

To verify the model described above, we performed a simulation with numerically-generated frequency noise with a 1/f noise spectrum and a fractional noise floor in the Allan deviation of $\sigma_0 = 4.4\times10^{-16}$~\cite{rosenband2012,hume_probing_2016}. For this simulation we used 300,000 clock interrogation cycles and varied the probe duration from 20 ms to 200 ms, with dead time equal to an average dead time between the $\Mg/\Al$ and $\Ca/\Al$ systems of 100 ms. At short averaging times, the behavior of the Allan deviation is determined by the flicker noise. At averaging times beyond 100 s, the simulation reaches a $1/\sqrt{\tau}$ slope as expected for the white spectrum of quantum projection noise. We fit a white noise model to the Allan deviation from the numerical simulation to find the asymptotic 1 second instability. Main text Fig.~3 shows the results of that simulation (assuming two identical clocks with uncorrelated noise) up to a maximum probe time of 0.18~s, beyond which fringe hops in the lock resulted in diverging instability.

The numerical estimate of the laser-limited instability is slightly above the analytical estimate because of non-linear error signal response at longer probe durations. Compared to typical clock operation, as described in~\cite{brewer_+_2019}, both of these estimates ignore details like probing multiple Zeeman states to track the magnetic field noise and probing from two opposing directions to monitor for drifts in the ion motion. As a result the laser-limited instability shown here is a best case scenario that can be compared to the instability observed in the correlation spectroscopy experiment. 

\section{Stability of correlation spectroscopy measurements}
\label{sec:correlation_stability}

To model the stability of correlation spectroscopy, we assume perfect $\pi/2$ pulses, but negligible laser coherence at the timescale of the Ramsey probe (i.e., $T_{\pi/2} \ll T_{\rm{coherence}} \ll T_{\rm R}$). We further assume the atomic coherence to be lifetime-limited. As described in the main text, the expression for the expectation value of the parity operator under these conditions is
\begin{equation}
    \label{avgparity}\tag{S9}
    \langle\hat{\Pi}\rangle = \frac12e^{-\Gamma T_\textrm{R}}\cos\left(\Delta_- T_\textrm{R} + \phi_-\right),
\end{equation}
where $\Gamma$ is the atomic excited state decay rate, $\Delta_-= \omega_{0,1}-\omega_{0,2}$ is the difference between atomic resonance frequencies and $\phi_-= \phi_{1}-\phi_{2}$ is a differential control phase. In a measurement of the relative frequency of two systems utilizing correlation spectroscopy the single-shot frequency uncertainty is given by
\begin{equation}
    \label{corr_freq_uncer}\tag{S10}
    \delta\Delta_- = \frac{\delta\Pi}{|d\langle\Pi\rangle/d\Delta_-|}.
\end{equation}
The measurement variance $(\delta\Pi)^2$ is limited by projection noise. Using Eq.~(\ref{avgparity}) and Eq.~(\ref{corr_freq_uncer}) we obtain an expression for the single-shot frequency uncertainty in correlation spectroscopy,
\begin{equation}\tag{S11}
    \delta\Delta_- = \frac2{T_{\rm R}}\frac{\sqrt{1 - \frac14 e^{-2\Gamma T_{\rm R}}\cos^2(\Delta_-T_{\rm R} + \phi_-)}}{|e^{-\Gamma T_{\rm R}}\sin(\Delta_-T_{\rm R}+\phi_-)|}.
\end{equation}
The single-shot measurement uncertainty can be minimized by probing on the maximum slope points of the correlation parity fringe, $\Delta_-T_{\rm R} + \phi_- = \pm \pi/2$, resulting in lifetime-limited measurement uncertainty of
\begin{equation}\tag{S12}
    \delta\Delta_- = \frac2{T_{\rm R}}e^{\Gamma T_{\rm R}}.
\end{equation} 
With this expression for the single-shot measurement uncertainty, the asymptotic measurement instability limited by projection-noise is given by
\begin{equation}
    \label{corr_instab}\tag{S13}
    \sigma_{\rm corr}(\tau) = \frac{\delta\Delta_-}{\omega_0}\sqrt{\frac{T_{\rm R}}{\tau}} = \frac{2}{\omega_0\sqrt{\tau T_{\rm R}}}e^{\Gamma T_{\rm R}},
\end{equation}
where we have used $\omega_{0,i} \approx \omega_0$. The probe duration that gives the lowest achievable measurement instability is $T_{\rm R, opt} = 1/(2\Gamma)$. The stability at $T_{\rm R, opt}$ is
\begin{equation}\tag{S14}
    \sigma_{\rm R,opt}(\tau) = \frac{2}{\omega_0}\sqrt{\frac{2e\Gamma}{\tau}}.
\end{equation}
For $\Al$ with $\omega_0 = 2\pi\times1.121\times10^{15}$ and $\Gamma = 1/20.6$~s, a value of $\sigma_{\rm R,opt}(\tau) = 1.5\times10^{-16}/\sqrt{\tau/s}$ is obtained.

\section{$\Ca/\Al$ Experimental Setup}
\label{sec:ca_al_clock}
%\todo[inline]{different species used but this might not be important since Ca is commonly used just give a rundown of which transitions serve what purpose in Calcium}
%\todo[inline]{should be just cite the Guggemos paper for info about Ca Al? }

The design of the $\Ca/\Al$ vacuum chamber and ion trap are similar to the $\Mg/\Al$ system described in Ref.~\cite{brewer_+_2019}. Sympathetic cooling and state detection are performed using $\Ca$, similar to methods described in Ref.~\cite{roos_quantum_1999}. One significant difference from Ref.~\cite{roos_quantum_1999} is that we perform EIT-assisted Doppler cooling which uses the dark resonance of the 397~nm and 866~nm transitions to increase the cooling rate and decrease the cooling limit~\cite{allcock_dark-resonance_2016}. 

A significant limitation to the up-time of the $\Mg/\Al$ system is collisions of the $\Al$ ion with molecular hydrogen background gas present in the vacuum chamber. Collisions with molecular hydrogen contributes to the systematic uncertainty through collisional heating and phase shifts~\cite{hankin_systematic_2019}, can cause excitation to unwanted meta-stable states such as ${}^3P_2$, and can cause Al-H$^+$ formation which is a significant limitation to the lifetime of the $\Al$ ion. To address this limitation, the $\Ca/\Al$ system uses a titanium vacuum chamber to reduce the hydrogen partial pressure relative to the $\Mg/\Al$ system which uses a stainless steel chamber. The external vacuum system components are made with grades II and V titanium \cite{minato_vacuum_1995,kurisu_titanium_2003} with the exception of a stainless steel flange on the ion pump. The reduction in the hydrogen partial pressure has reduced the rate of $\rm{Al-H^+}$ formation and unwanted meta-stable state excitation in the $\Ca/\Al$ system roughly by a factor of 2-3 as compared to the $\Mg/\Al$ system, although these rates have not been rigorously quantified.

\section{Optical pumping to inner Zeeman states of the $^1S_0$ manifold}
\label{sec:optical_pumping}
\begin{figure}
\includegraphics[scale = .4]{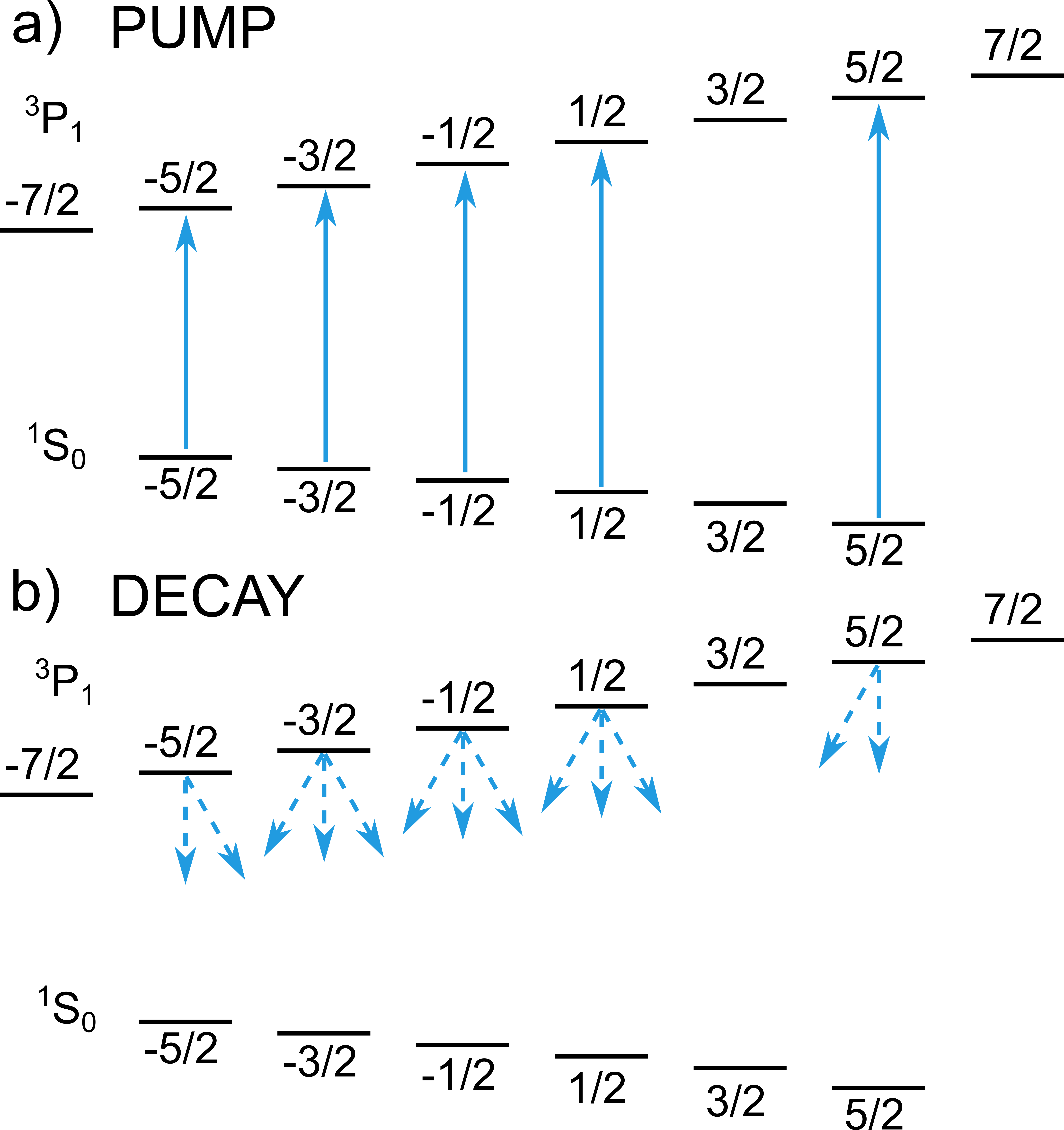}
\caption{\label{fig:3p1_pump}Diagram of the (a) pumping and the (b) decay for a single cycle of the optical pumping to prepare the $\ket{^1S_0,m_F=3/2}$ state. 12 cycles of this pumping is used to achieve a transition probability $\geq95\%$ in the observed Rabi lineshape for the $\ket{^1S_0,m_F = 3/2}\leftrightarrow\ket{^3P_0,m_F=1/2}$ transition. To ensure that the excited state has decayed we wait $300\rm~\mu$s before applying another set of $\pi$-pulses. }
\end{figure}

To reduce the sensitivity of the experiment to magnetic field noise we prepare the $\Al$ atom in each of the two atomic clocks into the $\ket{^1S_0,m_F = 3/2}$ Zeeman state so we can drive the $\ket{^1S_0,m_F = 3/2}\leftrightarrow\ket{^3P_0,m_F = 1/2}$ clock transition. This state preparation is done by optical pumping of $\Al$ on the $\ket{^1S_0}\rightarrow\ket{^3P_1}$ transition. To drive the population to the target state, a series of on-resonance $\pi$-polarized laser pulses are applied to all Zeeman sub-levels within $\ket{^1S_0}$, aside from the target state. A diagram of the pumping and decay cycles of the excited $\ket{^3P_1}$ state can be seen in Fig.~S\ref{fig:3p1_pump}. The frequency of these transitions, as mentioned in the text, are spaced $\approx$1~MHz apart from the nearest neighbor Zeeman transition with a quantization field of 1.5 to 1.7~G, and these optical pumping transitions (pulse durations $t_{\pi} > 50$~$\rm \mu s $) are frequency-resolved. After each series of 5 $\pi$-pulses we wait 300~$\rm \mu s$ for the ${}^3P_1$ state to decay ($^3P_1$ lifetime $\approx 300 \mu$s) before another cycle of pumping is applied. Experimentally, we determined that 12 cycles of optical pumping saturates the contrast in the $\ket{^1S_0,m_F = 3/2}\leftrightarrow\ket{^3P_0,m_F=1/2}$ transition. This pumping procedure can be used to prepare the state of $\Al$ into any of the $^1S_0$ Zeeman ground states.

%% Start here

\section{Differential noise sources}
\label{sec:differential}

The primary sources of differential noise that limit the atom-atom coherence between the two systems are magnetic field noise and variations in the clock laser optical path length between the two systems. In the following sections we describe methods used to reduce the effects of these differential noise sources.

\begin{figure*}[ht]
\includegraphics[scale = .70]{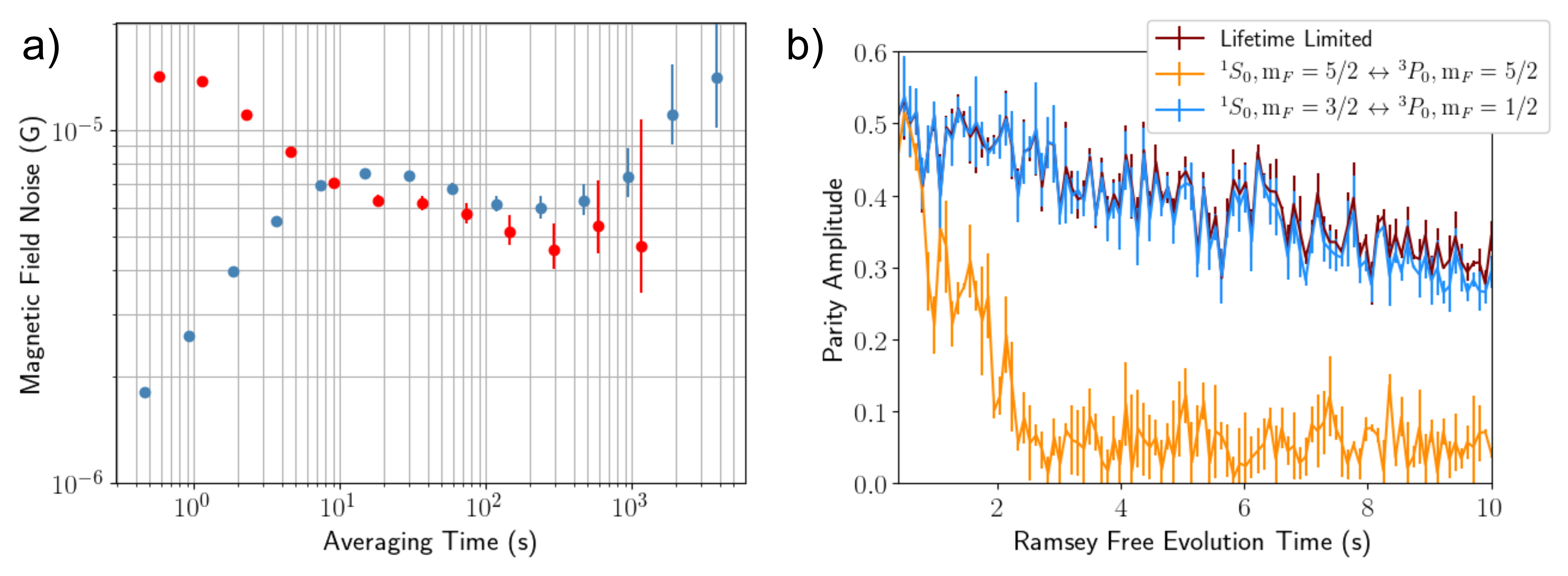}
\caption{\label{fig:mag_noise}Allan deviation of the measured magnetic field amplitude along the quantization axis, simultaneously sampled in the $\Mg/\Al$ system (red) and in the $\Ca/\Al$ system (blue). b) Simulated parity amplitude vs. Ramsey free evolution time plots using random laser phase noise and the measured differential magnetic field noise. Here, the red trend is neglecting the effects of the differential magnetic field noise and is only taking into account the lifetime limit of the excited state while the dark orange and light blue includes differential magnetic phase noise determined using the magnetic field sensitivity of the $\ket{^1S_0,m_F = 5/2}\leftrightarrow\ket{^3P_0,m_F = 5/2 }$ and $\ket{^1S_0,m_F = 3/2}\leftrightarrow\ket{^3P_0,m_F = 1/2 }$ transitions respectively. When probing the $\ket{^1S_0,m_F = 3/2}\leftrightarrow\ket{^3P_0,m_F = 1/2 }$ transition we expect to measure a parity amplitude consistent with the parity amplitude due to the lifetime limit.}
\end{figure*}

\subsection{Magnetic field stability}
In the main text we present data that utilized two different Zeeman levels in the $\ket{^1S_0}$ and $\ket{^3P_0}$ manifolds as the lower and upper states of the clock transition. We model the effect of differential first order Zeeman shifts between the ground and excited state of the clock transition due to magnetic field noise using
\begin{equation}\tag{S15}
\Delta f_{\rm Al}(B) = \mu_{\rm B} B(g_\textrm{P} m_F(^3P_0) - g_\textrm{S} m_F(^1S_0)),
\end{equation}
where $g_{\rm P} =-0.00197686(21)$, $g_{\rm S} = -0.00079248(14)$, $\mu_{\rm B}$ is the Bohr magneton, and $B$ is the instantaneous applied magnetic field~\cite{rosenband_observation_2007}. Initially, measurements were done using the $\ket{^1S_0,m_F=5/2}\leftrightarrow\ket{^3P_0,m_F=5/2}$ transition (Fig.~3 in the main text) which has a factor of 15 larger sensitivity to magnetic fields. When performing correlation spectroscopy on $\ket{^1S_0,m_F=5/2}\leftrightarrow\ket{^3P_0,m_F=5/2}$ we observe an optimum measurement stability at $\approx$~2~s due to the differential magnetic field noise between the two systems limiting the atom-atom coherence. 

To measure the magnetic field noise present in our systems we use the logic ion as a sensor. In the $\Mg$ system we lock a frequency-doubled DDS source to the $\ket{^2S_{1/2}, \rm{F} = 3, m_F = 3}\leftrightarrow\ket{^2S_{1/2}, \rm{F} = 3, m_F = 2}$ microwave transition that has a sensitivity to magnetic fields of 2.3 MHz/G and monitor the drift in the frequency to infer the change in the magnetic field as a function of time. In the $\Ca$ system we use Ramsey spectroscopy on a superposition of the $\ket{^2D_{5/2}, m_F = -5/2}$ and $\ket{^2D_{5/2}, m_F = 3/2}$ Zeeman states to increase the sensitivity of our magnetic field measurement. Using this superposition state we obtain a sensitivity to magnetic fields of 6.72 MHz/G. This superposition state is generated by first driving a $\pi/2$-pulse on the $\ket{^2S_{1/2},m_F = -1/2}\rightarrow\ket{^2D_{5/2}, m_F = -5/2}$ transition then a $\pi$-pulse on the $\ket{^2S_{1/2},m_F = -1/2}\rightarrow\ket{^2D_{5/2}, m_F = 3/2}$ transition. By locking the phase of the second $\pi/2$-pulse to the peak of the Ramsey fringe we can track changes in the magnetic field by converting the determined phase correction into a magnetic field.

\begin{figure*}[ht]
\includegraphics{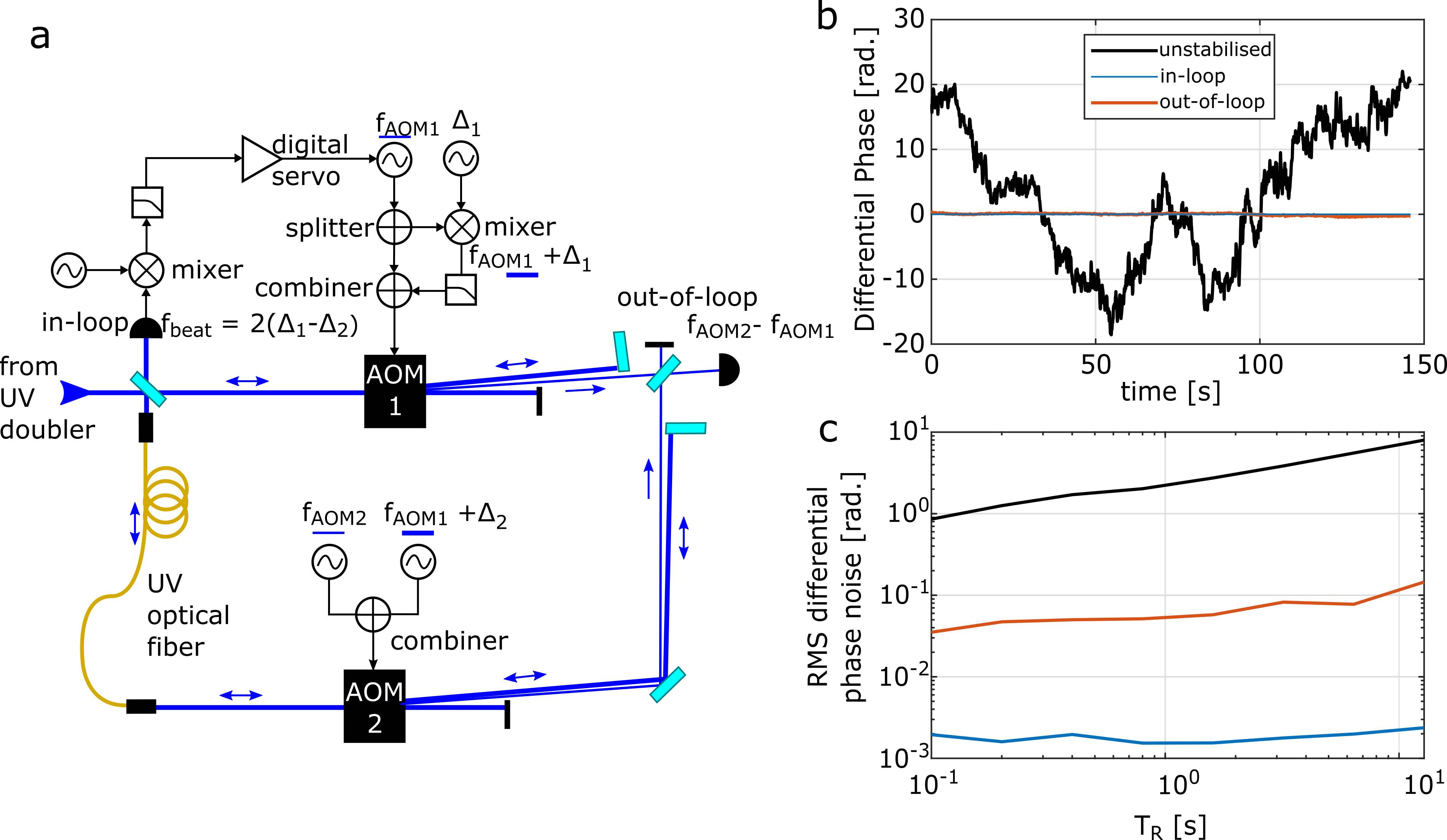}
\caption{\label{fig:path_length}Measurement of differential path length noise in a test setup. a) Diagram of the optical and electronic system for stabilizing and measuring differential phase noise between the two beam paths. This test setup effectively replaces the ions with a beam-splitter that allow us to form an out-of-loop beatnote for which the phase noise should approximate the phase noise of the real experiment (AOM: acousto-optic modulator). b) Time-series phase measurement of the out-of-loop and in-loop detectors (measured simultaneously) and the unstabilized phase noise observed on the out-of-loop detector immediately before the other measurements. c) Root-mean-squared phase fluctuations for a simulated Ramsey experiment using the three time-series measurements from b). }
\end{figure*}

To determine how the fringe contrast in the correlation spectroscopy experiment is affected by magnetic field noise, we perform numerical simulations of the experiment incorporating magnetic field noise from a simultaneous measurement of the magnetic field local to both experiments. By using this data in the simulation we can determine the effect that this representative magnetic field noise has on the parity amplitude. To simulate the parity fringes we begin with the single atom $\hat{\sigma}_{z,i}$ observable where $\langle\hat{\sigma}_{z,i}\rangle~=~\langle\cos(\phi_{\rm L}-\phi_{\textrm{diff},i}-\phi_i)\rangle$ and $i$ labels which atomic clock is being measured. In this equation $\phi_{\rm L}$ is the laser phase noise common to both systems, $\phi_{\textrm{diff},i}$ is the differential phase noise present in system $i$, and $\phi_i$ is used to scan the phase in one system and held constant in another to scan the relative phase between the two systems. The probe time is assumed to be much longer than the laser coherence time such that the laser phase can be modeled as a uniformly-distributed random number on the interval $\phi_{\rm L} \in [0,2\pi)$. The time series of the measured magnetic fields of the two systems is converted into a phase shift, which is inserted into the model as $\phi_{\textrm{diff},i}$. The simulated $\langle\hat{\sigma}_{z,i}\rangle$ time series for each system is then used to determine the parity $\langle\hat{\Pi}\rangle = \langle\hat{\sigma}_{z,1}\rangle\langle\hat{\sigma}_{z,2}\rangle$ which is averaged for each value of ($\phi_1-\phi_2$) and a fringe is fit to extract the contrast. Because the phase noise grows with $T_{\rm R}$ the effect of this measured magnetic field noise becomes more significant at longer Ramsey free evolution times as seen in Fig.~S\ref{fig:mag_noise}. This simulation suggests that switching to interrogating the inner manifold transition ($\ket{\ce{^1\mathit{S}_0},m_F = 3/2}\leftrightarrow\ket{ {}^3P_0,m_F=1/2}$) is necessary to achieve lifetime limited contrast of the parity fringe. By probing on the most magnetically insensitive transition in the Zeeman manifolds of $^1S_0$ and $^3P_0$ we are able to reach the lifetime limit of the clock transition.

\subsection{Optical path length noise}

 To minimize fluctuations in the laser beam optical path length, we actively control the differential phase of the light delivered to the two systems. Before implementing the optical path length stabilization described in the main text, we built a separate test setup that allowed us to measure the out-of-loop stability. A diagram of this setup is shown in Fig.~S\ref{fig:path_length}. The out-of-loop beatnote is generated by overlapping two laser beams, equivalent to the clock probe beams in the experiment, on a beam splitter. The phase of this beatnote includes noise contributions from any optical paths that are not common to the clock probe beams and their respective retroreflected beams. Here and in the experiment, the main contributions are the short air paths after AOM-1 and AOM-2. This measurement also includes any electronic noise introduced by either the phase-locked-loop or the out-of-loop phase measurement.
%The test setup is different from the actual experiment in a few details that we don't expect to change the performance significantly.  First there is only one AOM in each of the optical pathways which requires that we use two rf frequencies applied to each of these two AOMs to produce the probe beams and the retroreflected beams respectively.  To do this we use an rf splitter, mixer and combiner on the signal going to AOM-1 as depicted in the diagram to provide equal phase shifts to both the probe and retroreflected beams.  Here, the two rf signals are separated by about 80 MHz, as opposed to the 300 MHz frequency of the switch AOMs in the experiment, which means the two optical paths are somewhat closer to each other spatially in the test set up.  Separate measurements of differential noise as a function of optical path separation did not show a significant effect at these length scales.  Also, the total fiber path length in this test setup is 2 m as opposed to about 8 m in the experiment. Since these paths are in-loop we do not expect the extra fiber length to cause significant additional noise.

In Fig.~S\ref{fig:path_length}(b) and (c) we show the results of the differential phase noise measurements, both in-loop and out-of-loop, with the latter case plotted both for the stabilized and free-running (unstabilized) cases. While the out-of-loop noise measured here is worse than the in-loop noise, indicating that there is differential noise after the two AOMs, the residual noise remains negligible ($<\pi/20$~rad) for all probe times considered here.

\section{Data Analysis}
\label{sec:data_analysis}
\subsection{Error analysis with non-parametric bootstrapping}
\label{sec:bootstrapping}
Following Ref.~\cite{efron_bootstrap_1992} we utilize non-parametric bootstrapping to estimate the uncertainty of the fits to our data. Using our experimentally determined binomial distributions for each phase, we randomly draw $n$ events, where $n$ is equal to the number of measurements at a specific phase. The numerically sampled data is then averaged to obtain the parity at each measured phase and these points are fit using the equation $C\sin{(\phi_--\phi_0)}$ where $C$ is the parity amplitude and $\phi_0$ is the phase offset of the fringe. This sequence of resampling and fitting is repeated $N = 10,000$ times and the fit parameters $C$ and $\phi_0$ for each fit are recorded. The mean and standard deviation of these fit parameters converge for large $N$ and are used in the results of Fig.~3 in the main text. For Fig.~3(b) in the main text, we use the standard deviation of the $\phi_0$ bootstrap distribution to estimate the instability $\sigma_{\rm upper}$ that includes the effect of deadtime in the experiment. Here, the deadtime includes the overhead from state preparation and cooling and the time lost due to collisions and synchronization errors (as much as $76\%$ of data is affected by these events at short probe times). To calculate the uncertainty in $\sigma_{\rm{upper}}$ we first calculate the variance of this parameter under the assumption that the data used to determine the measurement standard deviation $s$ is normally distributed and $(n_{\rm{df}}-1)s^2/\sigma^2$ is a variable distributed as $\chi^2_{n_{\rm{df}}-1}$. In this analysis $\sigma^2$ is the true variance of the distribution and $n_{\rm{df}}$ is the degrees of freedom. The number of degrees of freedom can be determined by $n_{\rm{df}} = (n_{\phi}-n_{\textrm{fit}})$ where $n_{\phi}$ is the number of phases probed on the fringe and $n_{\textrm{fit}}$ is the number of fit parameters. 

\subsection{Factors limiting parity fringe contrast}
\label{sec:amplitude_limitation}
From fitting to the $\ket{^1S_0,m_F=3/2}\leftrightarrow\ket{^3P_0,m_F=1/2}$ experimental data in Fig.~3(a) we observe that the decay rate of the parity contrast as a function of Ramsey free evolution time is consistent with the lifetime of the ${}^3P_0$ excited state. However, in the same fit we observe a $T_{\rm{R}}=0$ parity fringe contrast of $A = 0.4\pm0.04$ where an ideal parity contrast would be $\rm A = 0.5$. Here, we discuss removal of data flagged in realtime and flagged in post processing and discuss effects which may be the cause of the reduction in the $T_{\rm R}=0$ parity fringe contrast.

Data which is flagged in real time consists of collision events and missed experiment triggers. Background gas collision events can be observed as either a loss in fluorescence from the cooling ion or as a change in the order of the two ions~\cite{hankin_systematic_2019}. Both of these signals are continuously monitored and collision events are filtered by removing the data point coincident with the event and the one immediately preceding. The total percentage of useful data was as small as 24$\%$ for short $T_\textrm{R}$ and as large as 71$\%$ for longer $T_\textrm{R}$. However, at short probe durations the loss of data is primarily a result of timing synchronization errors. Filtering of the data significantly reduces the mean duty-cycle for probing the clock transitions, but improves the contrast of the parity signal.

Separately, some collision events result in $\rm Al-H^+$ molecule formation. This can only occur when $\Al$ is in its ${}^3P_0$ excited state and a $\rm H_2$ molecule collides with energy greater than the reaction barrier needed to form $\rm Al-H^+$. These collision events can also result in the ion being excited to a metastable internal state that is not addressable by any of the experimental pulses. The likely metastable states for $\Al$ are $\ket{^3P_2}$ ($\tau=298.5$~s) and any Zeeman sublevel of $\ket{^3P_0}$ not addressed by the clock probe. Both of these types of collision events are filtered in post processing by checking for long measurement periods where no transitions are detected on one of the clocks.

\begin{table}
    \centering
    \begin{tabular}{|p{2.5cm}||p{1.75cm}|p{1.75cm}|}
        \hline
        \multicolumn{3}{|c|}{Data Filtering}\\
        \hline
        Ramsey~Free~Evolution~Time (s) & Total $\#$ of meas. & $\%$ data used\\
        \hline
        0.5 & 2224 & 24  \\
        \hline
        1.0 & 1035 & 71  \\
        \hline
        1.5 & 1317 & 50  \\
        \hline
        2.0 & 944 & 42  \\
        \hline
        3.0 & 706 & 66  \\
        \hline
        4.0 & 1040 & 63  \\
        \hline
        6.0 & 594 & 44  \\
        \hline
        8.0 & 791 & 43  \\
        \hline
    \end{tabular}
    \caption{Table detailing the number of measurements taken at each Ramsey free evolution time and the percent of data remaining following filtering. Total $\#$ of meas. is the total length of the data array before filtering and $\%$ data used is the data remaining following filtering for data affected by collisions and asynchronous probes. At short probe durations the loss of data is primarily a result of timing synchronization errors. }
\end{table}

%The data contributing to Fig.~M\ref{fig:combined_fringe} have been filtered for several effects, all of which are monitored in real time. These include synchronization errors, collisions with background gas molecules~\cite{hankin_systematic_2019}, and events in which at least one of the $\Al$ ions is excited, via collisions, to a metastable internal state that is not addressable by laser pulses.  The likely metastable states for $\Al$ are $\ket{^3P_2}$ ($\tau=298.5$~s) and any Zeeman sublevel of $\ket{^3P_0}$ that is not addressed by the clock probe. 

During the course of this measurement we discovered significant jitter and drifts in the CPU clocks used to match data measured on the two systems. To mitigate these timing errors, the CPU clock for the $\Mg/\Al$ system was used as an NTP server for the computer clock on the $\Ca/\Al$ system. Nevertheless, the final data could include some instances of asynchronous probes on the two systems, particularly at the shorter probe durations

When a photo-chemical reaction occurs and an $\rm Al-H^+$ molecule is formed, a new $\Al$ ion must be loaded. Following reloading, we observe no change in the $\ket{^1S_0}\leftrightarrow\ket{^3P_0}$ transition frequency. Because of this we are able to combine data from sequential runs at the same Ramsey free evolution time.

The parity amplitude $A = 0.4\pm0.04$ observed at short $T_{\rm{R}}$ is likely limited by imperfect state preparation and $\ket{^1S_0,m_F=3/2}\leftrightarrow\ket{^3P_0,m_F=1/2}$ $\pi$-pulse infidelity. Experimental optimization of the state preparation sequence using Rabi spectroscopy results in $5\%-10\%$ state preparation infidelity on a single system and we observe $\approx5\%$ $\pi$-pulse infidelity. This results in a $10\%-15\%$ reduction in the maximum achievable contrast when performing correlation spectroscopy on two clocks. One possible cause for the reduction in parity fringe contrast in longer datasets (e.g. 4~s) is long term drifts in the ${}^3P_1$ laser frequency. This is currently addressed by periodic recalibration of the ${}^3P_1$ lasers.

\bibliography{supp_biblo}